# Distinct moiré textures of in-plane electric polarizations for distinguishing moiré origins in homobilayers

Hongyi Yu[1,2,†], Ziheng Zhou[1], Wang Yao[3,4]

[1] Guangdong Provincial Key Laboratory of Quantum Metrology and Sensing & School of Physics and Astronomy, Sun Yat-Sen University (Zhuhai Campus), Zhuhai 519082, China
[2] State Key Laboratory of Optoelectronic Materials and Technologies, Sun Yat-Sen University (Guangzhou Campus), Guangzhou 510275, China
[3] Department of Physics, The University of Hong Kong, Hong Kong, China
[4] HKU-UCAS Joint Institute of Theoretical and Computational Physics at Hong Kong, China
[†] yuhy33@mail.sysu.edu.cn

**Abstract:** In binary compound 2D insulators/semiconductors such as hexagonal boron nitride (hBN), the different electron affinities of atoms can give rise to out-of-plane electric polarizations across inversion asymmetric van der Waals interface of near $0°$ interlayer twisting. Here we show that at a general stacking order where sliding breaks $2\pi/3$-rotational symmetry, the interfacial charge redistribution also leads to an in-plane electric polarization, with a magnitude comparable to that of the out-of-plane ones. The effect is demonstrated in hBN bilayers, as well as in biased graphene bilayers with gate-controlled interlayer charge redistributions. In long wavelength moiré patterns, the in-plane electric polarizations determined by the local interlayer stacking registries constitute topologically nontrivial spatial textures. We show that these textures can be used to distinguish moiré patterns of different origins from twisting, biaxial- and uniaxial-heterostrain, where vector fields of electric polarizations feature Bloch-type merons, Néel-type merons, and anti-merons, respectively. Combinations of twisting and heterostrain can further be exploited for engineering various electric polarization textures including 1D quasiperiodic lattices.

**Keywords:** moiré patterns, van der Waals stacking, electric polarization, hexagonal boron nitride, graphene

**PACS:** 73.21.Cd, 73.21.Ac, 77.80.–e, 73.43.Cd

Long-wavelength moiré patterns formed in van der Waals layered insulator/semiconductor systems have emerged as a fascinating platform for exploring novel physical phenomena [1,2], including moiré excitons [3-8], and electron correlation phenomena [9-20] where many-body interaction becomes significant due to the flat dispersion of minibands [21-23]. The wavelength of the moiré pattern usually lies between several to several tens nm, much larger than the monolayer lattice constant. In a nanoscale region small compared to the moiré supercell, the interlayer coupling is determined by the local stacking registry, which varies smoothly and periodically in the long-wavelength moiré landscape. This gives rise to spatial modulations in a variety of physical quantities including the interlayer distance, local bandgap, optical transition dipole, and magnetization [24-28].

Some less intuitive moiré modulated phenomena are recently noted in the context of binary compound homobilayer insulators/semiconductors such as hexagonal boron nitride (hBN) and semiconducting transition-metal dichalcogenides (TMDs). With the two layers being identical, layer pseudospin describing the layer occupation of the carrier becomes an active quantum degree of freedom, which is associated with an out-of-plane electric polarization. Theoretical studies have shown that in the moiré landscape the layer pseudospin is subject to an effective Zeeman field of topologically nontrivial spatial texture, underlying various topological phenomena in homobilayer moiré [29-33]. In particular, the out-of-plane component of such pseudospin Zeeman field arises from a spontaneous interfacial electric polarization that is determined by the stacking registries between the layers [34-41]. Taking AB/BA-stacked hBN bilayers as an example, where B atoms of one layer are vertically aligned with N atoms of the other, it has been pointed out that different electron affinities of the atoms lead to a small amount of electron transfer from N to nearest B atoms across the van der Waals interface [42]. The resultant out-of-plane electric polarization has opposite signs in the AB and BA configurations, which are connected by an interlayer sliding. The observation of the out-of-plane electric polarization and its sliding control have been reported in commensurate and marginally twisted bilayer hBN and TMDs [34-41], which can be exploited for ferroelectric functionalities, as well as for noninvasive engineering of superlattice potentials in adjacent 2D materials [43].

Here we show that the nature of the stacking registry determined electric polarization necessarily implies the presence of an in-plane component and its moiré-patterned spatial texture, accompanying the out-of-plane one already observed. Just like the AB/BA stacking where $2\pi/3$-rotational ($\hat{C}_3$) symmetry only allows an out-of-plane electric polarization, there exist stacking registries with $\pi$-rotational ($\hat{C}_2$) symmetry about an in-plane axis which only allows an in-plane polarization [44]. Long wavelength moiré patterns formed by twisting, biaxial- or uniaxial-heterostrain have distinct spatial distributions of the $\hat{C}_2$-symmetric locales. While this difference has been neglected in other contexts of moiré phenomena, we show that it dictates distinct vector fields of the polarization textures in moiré patterns of the three different origins, corresponding to superlattices of Bloch-type merons, Néel-type merons and anti-merons, respectively. Combinations of twisting and heterostrain can further be exploited for engineering various polarization textures including 1D quasiperiodic lattices. The effect is demonstrated in bilayer hBN of the R-type stacking, as well as in biased graphene bilayer and H-type hBN bilayer with the gate-controlled interlayer charge redistribution. We find that the in-plane electric polarization at the $\hat{C}_2$-symmetric locale can have comparable magnitude to that of the well characterized sizable out-of-plane ones at $\hat{C}_3$-symmetric locales.

We first give an intuitive understanding to the out-of-plane and in-plane electric polarizations in van der Waals bilayer semiconductor/insulator systems induced by the interlayer coupling. A quantitative calculation follows. We use hBN as an illustration because of its simple band structure, while the generalization to other systems like TMDs [37-40,44] is straightforward. In bilayer hBN, a small amount of the electrons in the N atom can be transferred to the nearby B atoms in the other layer due to their different electron affinities [42]. Such an interlayer charge redistribution gives rise to local electric dipoles pointing from B to N, and the overall polarization determined by the summation of them is generally finite (see Fig. 1(a)). We consider an R-type bilayer hBN with a $0°$ interlayer twisting, characterized by an interlayer translation $\mathbf{r}_0$. Fig. 1(b) shows three high-symmetry stacking orders. The top and middle panels correspond to the so-called AB and BA stackings with $\mathbf{r}_0 = \mathbf{a}_D/3$ and $2\mathbf{a}_D/3$, respectively [45,46], where $\mathbf{a}_D \equiv \mathbf{a}_1 + \mathbf{a}_2$ is the long diagonal line of the unit cell. These stackings are $\hat{C}_3$-symmetric since a B atom in one layer horizontally overlaps with an N atom in the other layer, where the interlayer charge redistribution results in electric dipoles along the out-of-plane direction. Because these stackings have the minimum interlayer distances between B and N, the magnitudes of the charge redistribution and the resultant out-of-plane electric polarization are expected to be the largest. The lower panel of Fig. 1(b) and its $\hat{C}_3$ rotations correspond to the intermediate stackings between AB and BA, denoted as IM. It has $\hat{C}_2$ symmetry about an axis within the 2D plane, allowing an electric dipole in-plane but not out-of-plane. In Fig. 1(b) we use green arrows to indicate local electric dipoles between nearest interlayer B-N pairs. For AA stacking with $\mathbf{r}_0 = 0$ (not shown), the bilayer structure has $\hat{C}_3$ and in-plane mirror symmetries, thus the total electric polarization vanishes. For general stacking patterns with arbitrary $\mathbf{r}_0$, the induced electric polarizations are expected to have both the out-of-plane and in-plane components (Fig. 1(a)).

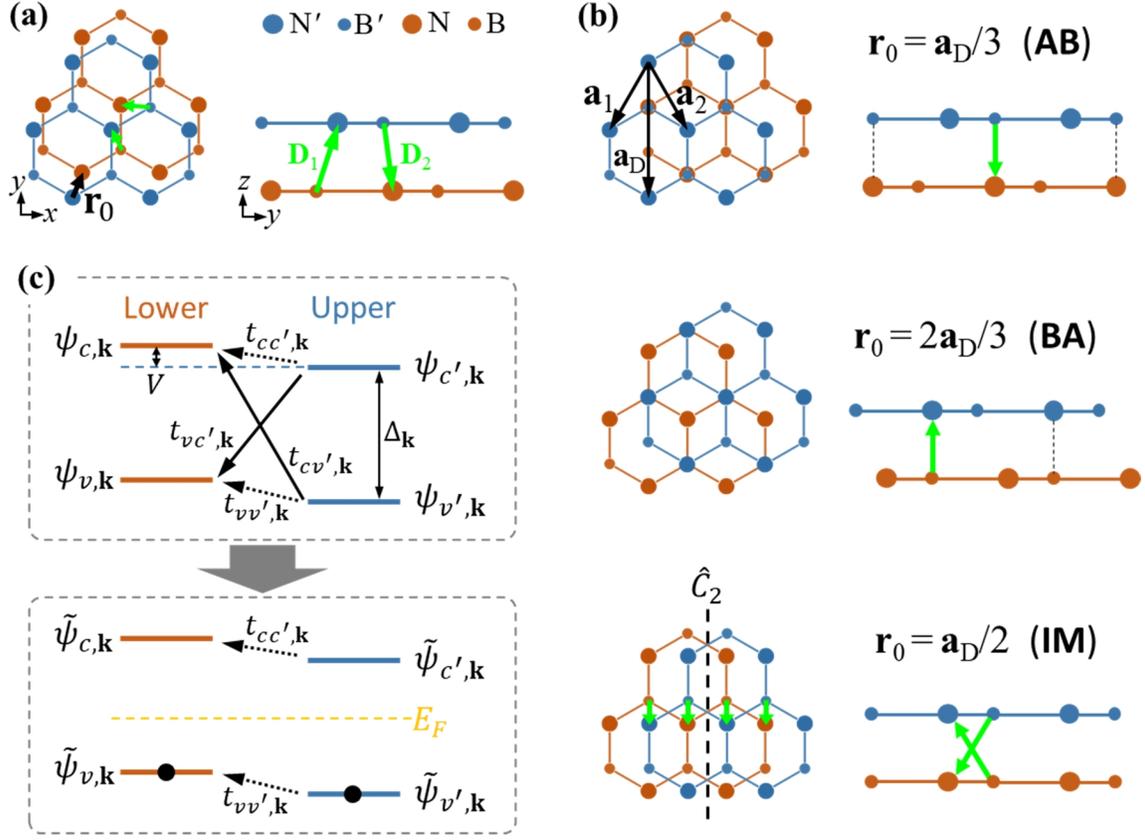

Fig. 1 (a) Top and side views of an R-type bilayer hBN with an arbitrary interlayer translation $\mathbf{r}_0$. The large (small) blue dots denote N (B) atoms in the upper layer, and the large (small) orange dots denote N (B) atoms in the lower layer. The green arrows indicate the local electric dipole pointing from a B atom in one layer to a nearest N atom in the other layer, induced by the interlayer charge redistribution. (b) The three high-symmetry stacking patterns AB, BA and IM (from up to down). AB and BA stackings are $\hat{C}_3$-symmetric with out-of-plane electric polarizations, whereas IM stacking with in-plane $\hat{C}_2$ symmetry has an in-plane electric polarization. $\mathbf{a}_D \equiv \mathbf{a}_1 + \mathbf{a}_2$ is the long diagonal line of the unit cell. (c) A schematic illustration of the interlayer couplings (black arrows) for the conduction and valence bands. $V$ is the interlayer bias. The magnitudes of $t_{vc',\mathbf{k}}$ and $t_{cv',\mathbf{k}}$ are much smaller than the energy separation $\Delta_\mathbf{k}$ thus can be eliminated by a second-order perturbation (see Eq. (1)). The Fermi level lies in the gap so both valence states are occupied whereas both conduction states are empty.

To quantitively calculate the induced electric polarization, we start from a layer-decoupled bilayer hBN. The lower-layer (upper-layer) Bloch state at wave vector $\mathbf{k}$ is denoted as $\psi_{n,\mathbf{k}}$ ($\psi_{n',\mathbf{k}}$) with an energy $E_{n,\mathbf{k}}$ ($E_{n',\mathbf{k}}$). Here we consider the two $\pi$-bands of hBN [45], termed as conduction and valence bands ($n = c, v$). The interlayer coupling is treated as a perturbation, which couples $\psi_{n,\mathbf{k}}$ and $\psi_{n',\mathbf{k}}$ with a strength $t_{nn',\mathbf{k}}$, see Fig. 1(c). As $|t_{nn',\mathbf{k}}|$ is much weaker than the band gap, the effect of cross-bandgap interlayer couplings $t_{vc',\mathbf{k}}$ and $t_{cv',\mathbf{k}}$ can be accounted by a second-order perturbation, which changes the valence bands in the two layers to

$$|\tilde{\psi}_{v,\mathbf{k}}\rangle \approx \left(1 - \frac{1}{2}\left|\frac{t_{vc',\mathbf{k}}}{D_\mathbf{k} - V}\right|^2\right)|\psi_{v,\mathbf{k}}\rangle - \frac{t^*_{vc',\mathbf{k}}}{D_\mathbf{k} - V}|\psi_{c',\mathbf{k}}\rangle,$$

$$|\tilde{\psi}_{v',\mathbf{k}}\rangle \approx \left(1 - \frac{1}{2}\left|\frac{t_{cv',\mathbf{k}}}{D_\mathbf{k} + V}\right|^2\right)|\psi_{v',\mathbf{k}}\rangle - \frac{t_{cv',\mathbf{k}}}{D_\mathbf{k} + V}|\psi_{c,\mathbf{k}}\rangle \quad (1)$$

We have used the notation $D_\mathbf{k} \equiv E_{c',\mathbf{k}} - E_{v,\mathbf{k}} \approx E_{c,\mathbf{k}} - E_{v',\mathbf{k}}$, and $V$ corresponds to an interlayer bias. The electric polarization can be calculated from the valence band eigenstates $f_{v_1,\mathbf{k}}$ and $f_{v_2,\mathbf{k}}$ of the coupled bilayer, which are the hybridizations of $\tilde{\psi}_{v,\mathbf{k}}$ and $\tilde{\psi}_{v',\mathbf{k}}$ by the interlayer coupling term $t_{vv',\mathbf{k}}$. However, we find the result is determined only by the perturbative effect of the cross-bandgap terms $t_{vc',\mathbf{k}}$ and $t_{cv',\mathbf{k}}$, independent on either $t_{vv',\mathbf{k}}$ or $t_{cc',\mathbf{k}}$. This is due to the fact that $\tilde{\psi}_{v,\mathbf{k}}$ and $\tilde{\psi}_{v',\mathbf{k}}$ are fully occupied (and the $t_{vc',\mathbf{k}}$- and $t_{cv',\mathbf{k}}$-perturbed conduction bands are all empty), so the hybridization induced by $t_{vv',\mathbf{k}}$ ($t_{cc',\mathbf{k}}$) does not introduce charge redistribution up to the second order. The electric polarization can then be calculated using $\tilde{\psi}_{v,\mathbf{k}}$ and $\tilde{\psi}_{v',\mathbf{k}}$ in Eq. (1) above.

Since the two layers are well separated by an interlayer distance $d \approx 0.33$ nm, one then defines the out-of-plane polarization operator $\hat{P}_z \equiv \frac{d}{2}\left(\sum_{n,\mathbf{k}}|\psi_{n,\mathbf{k}}\rangle\langle\psi_{n,\mathbf{k}}| - \sum_{n',\mathbf{k}}|\psi_{n',\mathbf{k}}\rangle\langle\psi_{n',\mathbf{k}}|\right)$, where we have set the elementary charge unit $e = 1$. The summation of $\hat{P}_z$ expectation values over all the occupied states $\tilde{\psi}_{v,\mathbf{k}}$ and $\tilde{\psi}_{v',\mathbf{k}}$ gives the out-of-plane polarization:

$$P_z = \frac{2d}{(2\pi)^2}\int d\mathbf{k}\left(\left|\frac{t_{cv',\mathbf{k}}}{D_\mathbf{k} + V}\right|^2 - \left|\frac{t_{vc',\mathbf{k}}}{D_\mathbf{k} - V}\right|^2\right) \quad (2)$$

The in-plane electric polarization is given by the summation of the Berry connections for the occupied states [44,47], i.e., $\mathbf{P}_\parallel = \frac{-2}{(2\pi)^2}\int d\mathbf{k}\left(i\langle\tilde{u}_{v,\mathbf{k}}|\frac{\partial\tilde{u}_{v,\mathbf{k}}}{\partial\mathbf{k}}\rangle + i\langle\tilde{u}_{v',\mathbf{k}}|\frac{\partial\tilde{u}_{v',\mathbf{k}}}{\partial\mathbf{k}}\rangle\right)$ with $\tilde{u}_{v/v',\mathbf{k}}$ the periodic part of $\tilde{\psi}_{v/v',\mathbf{k}}$. A straightforward derivation results in

$$\mathbf{P}_{\|} = \frac{-2}{(2\pi)^2}\int d\mathbf{k}\,(\mathbf{A}_{v,\mathbf{k}} + \mathbf{A}_{v',\mathbf{k}}) + \frac{2}{(2\pi)^2}\int d\mathbf{k}\,\Delta\mathbf{p}_{\|,\mathbf{k}}, \tag{3}$$

$$\Delta\mathbf{p}_{\|,\mathbf{k}} \equiv \mathrm{Im}\left(\frac{t_{vc',\mathbf{k}}}{(D_\mathbf{k}-V)^2}\frac{\partial t^*_{vc',\mathbf{k}}}{\partial \mathbf{k}} + \frac{t^*_{cv',\mathbf{k}}}{(D_\mathbf{k}+V)^2}\frac{\partial t_{cv',\mathbf{k}}}{\partial \mathbf{k}}\right)$$
$$+ \left|\frac{t_{vc',\mathbf{k}}}{D_\mathbf{k}-V}\right|^2 (\mathbf{A}_{v,\mathbf{k}} - \mathbf{A}_{c',\mathbf{k}}) + \left|\frac{t_{cv',\mathbf{k}}}{D_\mathbf{k}+V}\right|^2 (\mathbf{A}_{v',\mathbf{k}} - \mathbf{A}_{c,\mathbf{k}}). \tag{4}$$

Here $\mathbf{A}_{n,\mathbf{k}} \equiv i\langle u_{n,\mathbf{k}}|\frac{\partial u_{n,\mathbf{k}}}{\partial \mathbf{k}}\rangle$ and $\mathbf{A}_{n',\mathbf{k}} \equiv i\langle u_{n',\mathbf{k}}|\frac{\partial u_{n',\mathbf{k}}}{\partial \mathbf{k}}\rangle$ are the Berry connections of the layer-decoupled bands. On the right-hand-side of Eq. (3), the first term corresponds to electric polarizations of individual layers, which can be finite when under a uniaxial strain. Below we drop the first term, and focus on the second term which originates purely from the cross-bandgap interlayer coupling $t_{vc',\mathbf{k}}$ and $t_{cv',\mathbf{k}}$.

To obtain $t_{cv',\mathbf{k}}$ and $t_{vc',\mathbf{k}}$ we use the minimal tight-binding model involving the $p_z$-orbitals of B and N atoms, combined with the interlayer coupling form under the two-center approximation [45,46]. The bilayer Hamiltonian can be written in the form

$$\hat{H} = \begin{pmatrix} D/2+V & h_\mathbf{k} & h_{\mathbf{AA'},\mathbf{k}} & h_{\mathbf{AB'},\mathbf{k}} \\ h^*_\mathbf{k} & -D/2+V & h_{\mathbf{BA'},\mathbf{k}} & h_{\mathbf{BB'},\mathbf{k}} \\ h^*_{\mathbf{AA'},\mathbf{k}} & h^*_{\mathbf{BA'},\mathbf{k}} & D'/2 & h_\mathbf{k} \\ h^*_{\mathbf{AB'},\mathbf{k}} & h^*_{\mathbf{BB'},\mathbf{k}} & h^*_\mathbf{k} & -D'/2 \end{pmatrix} \tag{5}$$

Here $\mathbf{A}$ and $\mathbf{B}$ ($\mathbf{A'}$ and $\mathbf{B'}$) denote the two sublattices in the lower (upper) layer, $D$ ($D'$) being their on-site energy difference. $h_\mathbf{k} = t(e^{i\mathbf{k}\cdot\Delta\mathbf{R}_1} + e^{i\mathbf{k}\cdot\Delta\mathbf{R}_2} + e^{i\mathbf{k}\cdot\Delta\mathbf{R}_3})$, with $t \approx 2.4$ eV the intralayer hopping strength between the nearest-neighbor $\mathbf{A}$ and $\mathbf{B}$ sites and $\Delta\mathbf{R}_{1,2,3}$ their displacements. The interlayer couplings have the forms $h_{\mathbf{AA'},\mathbf{k}} = \sum_\mathbf{G} t_{\mathbf{AA'}}(\mathbf{k}+\mathbf{G})e^{i\mathbf{G}\cdot\mathbf{r}_0}$, $h_{\mathbf{BB'},\mathbf{k}} = \sum_\mathbf{G} t_{\mathbf{BB'}}(\mathbf{k}+\mathbf{G})e^{i\mathbf{G}\cdot\mathbf{r}_0}$, $h_{\mathbf{AB'},\mathbf{k}} = \sum_\mathbf{G} t_{\mathbf{AB'}}(\mathbf{k}+\mathbf{G})e^{i\mathbf{G}\cdot(\mathbf{r}_0-\Delta\mathbf{R}_1)}$ and $h_{\mathbf{BA'},\mathbf{k}} = \sum_\mathbf{G} t_{\mathbf{BA'}}(\mathbf{k}+\mathbf{G})e^{i\mathbf{G}\cdot(\mathbf{r}_0+\Delta\mathbf{R}_1)}$, with $\mathbf{G}$ the reciprocal lattice vector. Here $t_{XY'}(\mathbf{k})$ is the Fourier transform of the interlayer hopping $t_{XY'}(\mathbf{r})$ between a $Y' = \mathbf{A'},\mathbf{B'}$ sublattice in the upper layer and an $X = \mathbf{A},\mathbf{B}$ sublattice in the lower layer separated by an in-plane displacement $\mathbf{r}$. We use the approximated form $t_{XY'}(\mathbf{k}) = t^0_{XY'} e^{-k^2\sigma^2/4}$ with $\sigma$ the decay length. We note that bilayer graphene is described by the same Hamiltonian with $D = D' = 0$ as $\mathbf{A}$, $\mathbf{A}'$, $\mathbf{B}$ and $\mathbf{B}'$ all correspond to carbon atoms.

We first consider the R-type bilayer hBN (i.e., $0°$ twisting), where $\mathbf{A}$ and $\mathbf{A}'$ ($\mathbf{B}$ and $\mathbf{B}'$) correspond to B (N) atoms thus $D' = D \approx 4.5\,\mathrm{eV}$. One then obtains

$$t_{vc',\mathbf{k}} = \sum_{\mathbf{G}} \left[ a_{\mathbf{k}} b_{\mathbf{k}} \left( t^0_{BB'} - t^0_{NN'} \right) + b_{\mathbf{k}}^2 t^0_{BN'} e^{-i\mathbf{G}\cdot\Delta\mathbf{R}_1} - a_{\mathbf{k}}^2 t^0_{NB'} e^{i\mathbf{G}\cdot\Delta\mathbf{R}_1} \right] e^{i\mathbf{G}\cdot\mathbf{r}_0 - \frac{(\mathbf{k}+\mathbf{G})^2 s^2}{4}},$$

$$t_{cv',\mathbf{k}} = \sum_{\mathbf{G}} \left[ a_{\mathbf{k}}^* b_{\mathbf{k}}^* \left( t^0_{BB'} - t^0_{NN'} \right) + b_{\mathbf{k}}^{*2} t^0_{NB'} e^{i\mathbf{G}\cdot\Delta\mathbf{R}_1} - a_{\mathbf{k}}^{*2} t^0_{BN'} e^{-i\mathbf{G}\cdot\Delta\mathbf{R}_1} \right] e^{i\mathbf{G}\cdot\mathbf{r}_0 - \frac{(\mathbf{k}+\mathbf{G})^2 s^2}{4}}, \quad (6)$$

$$\mathbf{A}_{v,\mathbf{k}} = -\mathbf{A}_{c,\mathbf{k}} = \mathbf{A}_{v',\mathbf{k}} = -\mathbf{A}_{c',\mathbf{k}} = i a_{\mathbf{k}} \frac{\partial a_{\mathbf{k}}^*}{\partial \mathbf{k}} + i b_{\mathbf{k}} \frac{\partial b_{\mathbf{k}}^*}{\partial \mathbf{k}}.$$

Here $a_{\mathbf{k}} = \dfrac{D/2 + \sqrt{D^2/4 + |h_{\mathbf{k}}|^2}}{\sqrt{|h_{\mathbf{k}}|^2 + \left(D/2 + \sqrt{D^2/4 + |h_{\mathbf{k}}|^2}\right)^2}}$ and $b_{\mathbf{k}} = \dfrac{h_{\mathbf{k}}^*}{\sqrt{|h_{\mathbf{k}}|^2 + \left(D/2 + \sqrt{D^2/4 + |h_{\mathbf{k}}|^2}\right)^2}}$. We set the parameters as $t^0_{AA'} = t^0_{BB'} = 0.8\,\mathrm{eV}$, $t^0_{BB'} = t^0_{NN'} = 0.5\,\mathrm{eV}$, $t^0_{AB'} = t^0_{BA'} = t^0_{BN'} = t^0_{NB'} = 0.6\,\mathrm{eV}$, $s = 0.15\,\mathrm{nm}$.

Fig. 2(a) shows $P_z$ as a function of $\mathbf{r}_0$ under $V = 0$, where $\mathbf{r}_0$ varies in a monolayer unit cell. We can see that $|P_z|$ reaches maximum at AB and BA stackings, and vanishes at lines connecting nearest-neighbor AA stackings. The maximum polarization corresponds to a sizable charge density of $|P_z|/d \sim 0.01\,\mathrm{nm}^{-2}$ in each layer, and this value agrees with the experimental observations [34-36] and First-principles calculations [43]. Fig. 2(b) shows $P_z$ line cuts for $\mathbf{r}_0$ along the long diagonal line $\mathbf{a}_D$ of the unit cell, where a large interlayer bias $V = 0.5\,\mathrm{eV}$ only results in slight changes compared to $V = 0$. Meanwhile we have calculated the $\mathbf{k}$-dependent quantity $\Delta r_{\mathbf{k}} \equiv \langle \psi_{v,\mathbf{k}} | \hat{P}_z | \psi_{v,\mathbf{k}} \rangle / d + \langle \psi_{v',\mathbf{k}} | \hat{P}_z | \psi_{v',\mathbf{k}} \rangle / d = \left( |t_{cv',\mathbf{k}}|^2 - |t_{vc',\mathbf{k}}|^2 \right) / D_{\mathbf{k}}^2$, which gives the amount of electron transferred from the lower- to upper-layer at wave vector $\mathbf{k}$. The value of $\Delta r_{\mathbf{k}}$ for BA stacking as a function of $\mathbf{k}$ is indicated in Fig. 2(c). Obviously, most of the transferred electrons come from $\pm \mathbf{K}$ valleys, while the contributions near $\Gamma$ are negligible.

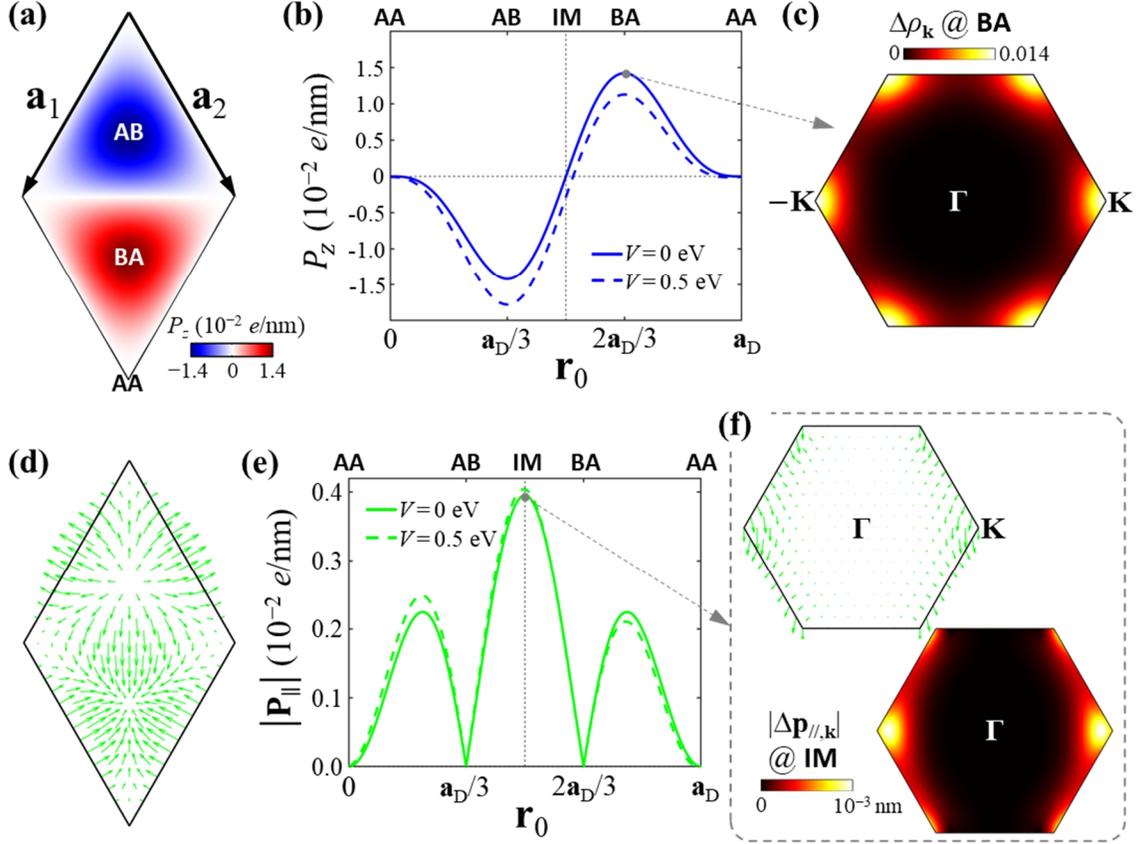

Fig. 2 (a) The out-of-plane polarization $P_z$ as a function of the interlayer translation $\mathbf{r}_0$ under $V=0$. $P_z$ has the largest magnitude at AB or BA stacking, and vanishes at lines connecting nearest-neighbor AA stackings. (b) The $P_z$ line cuts along the long diagonal line $\mathbf{a}_D$, for $V=0$ and 0.5 eV. (c) The $\mathbf{k}$-space distribution of $\Delta\rho_\mathbf{k}$ for BA stacking, which gives the amount of interlayer charge redistribution at wave vector $\mathbf{k}$. (d) The direction of the in-plane polarization $\mathbf{P}_\parallel$ as a function of $\mathbf{r}_0$ under $V=0$. The arrow length corresponds to $|\mathbf{P}_\parallel|$. (e) The $|\mathbf{P}_\parallel|$ line cuts along $\mathbf{a}_D$ for $V=0$ and 0.5 eV. (f) The direction (upper panel) and magnitude (lower panel) of the in-plane electric polarization $\Delta\mathbf{p}_{\parallel,\mathbf{k}}$ as functions of $\mathbf{k}$ in the full Brillouin zone, for an IM stacking.

For $\mathbf{P}_\parallel$ under $V=0$, we show its direction as a function of $\mathbf{r}_0$ in Fig. 2(d), where the arrow length corresponds to the magnitude. For a better illustration, in Fig. 2(e) we also show $|\mathbf{P}_\parallel|$ line cuts for $\mathbf{r}_0$ along $\mathbf{a}_D$, which is barely affected by $V$. $\mathbf{P}_\parallel$ vanishes for the three $\hat{C}_3$-symmetric stackings (AA, AB and BA), and reaches maximum for the three IM stackings where $P_z$ vanishes. Remarkably, the peak value of $\mathbf{P}_\parallel$ at IM stackings is of the same order of magnitude to that of $P_z$ at AB/BA stacking. It is worth pointing out that unlike the Berry connection $\mathbf{A}_{v/v',\mathbf{k}}$ that is gauge-dependent at a given **k** but becomes gauge-invariant after summing over the entire Brillouin zone, the quantity $\Delta \mathbf{p}_{\parallel,\mathbf{k}}$ given in Eq. (4) is gauge-invariant for every single **k**, which corresponds to the in-plane polarization contributed by occupied states at **k**. In Fig. 2(f), we show our calculated directions and magnitudes of $\Delta \mathbf{p}_{\parallel,\mathbf{k}}$ as functions of **k** for an IM stacking. Again, states near $\pm \mathbf{K}$ contribute the most, while those near $\Gamma$ have negligible contributions.

Now we consider the H-type bilayer hBN, i.e., $60°$ twisting, where $\mathbb{A}$ and $\mathbb{B}'$ ($\mathbb{A}$ and $\mathbb{B}'$) correspond to B (N) atoms and $\Delta = -\Delta' \approx 4.5\,\text{eV}$. Note that the system is inversion-symmetric at $V=0$, so $P_z(V=0)$ and $\mathbf{P}_\parallel(V=0)$ must vanish and $P_z \approx \left.\frac{\partial P_z}{\partial V}\right|_{V\to 0} V$ and $\mathbf{P}_\parallel \approx \left.\frac{\partial \mathbf{P}_\parallel}{\partial V}\right|_{V\to 0} V$ for $V = \Delta$. Using the same parameters as the R-type hBN, we obtain $\left.\frac{\partial P_z}{\partial V}\right|_{V\to 0}$ as shown in Fig. 3(a). At small $V\ (=\Delta)$, the magnitude of $P_z$ varies dramatically with the interlayer stacking registry $\mathbf{r}_0$ (Fig. 3(a)), whereas the sign is always opposite to $V$. $\mathbf{P}_\parallel$, on the other hand, can be calculated from its derivative that has a band geometric origin as [44]

$$\frac{\partial \mathbf{P}_\parallel}{\partial \lambda} = \frac{-1}{\rho^2}\oint d\mathbf{k}\left(\mathbf{W}^{(v_1)}_{\mathbf{k},\lambda} + \mathbf{W}^{(v_2)}_{\mathbf{k},\lambda}\right). \tag{7}$$

Here $\mathbf{W}^{(m)}_{\mathbf{k},\lambda} \equiv \sum_{n\neq m} \text{Im}\left(\langle f_{m,\mathbf{k}}|\frac{\partial \hat{H}}{\partial \mathbf{k}}|f_{n,\mathbf{k}}\rangle\langle f_{n,\mathbf{k}}|\frac{\partial \hat{H}}{\partial \lambda}|f_{m,\mathbf{k}}\rangle\right)/(E_{m,\mathbf{k}} - E_{n,\mathbf{k}})^2$ is the Berry curvature in $(\mathbf{k},\lambda)$ space, $\lambda$ denoting $V$ and $\mathbf{r}_0 \equiv (x_0, y_0)$, which parameterize the Hamiltonian $\hat{H}$ in Eq. (5). $m, n = v_1, v_2, c_1, c_2$ is the band index, and $f_{v_{1,2},\mathbf{k}}$ ($f_{c_{1,2},\mathbf{k}}$) denotes Bloch states of the layer-hybridized valence (conduction) bands. Fig. 3(b) and 3(c) show the direction and magnitude of $\left.\frac{\partial \mathbf{P}_\parallel}{\partial V}\right|_{V\to 0}$, respectively. Besides the $\hat{C}_3$-symmetric stackings AA', AB' and A'B with $+1$ winding numbers (red dashes circles in Fig. 3(b)), $\mathbf{P}_\parallel$ also vanishes at three low-symmetry stackings where the winding numbers are $-1$ (blue dashes circles in Fig. 3(b)). Here the winding number is defined as the number of counterclockwise rotations of $\mathbf{P}_\parallel$ when the path circles the given point once [48]. We note that the strengths of $P_z$ and $\mathbf{P}_\parallel$ in the H-type bilayer hBN are much weaker than in the R-type case when $\Delta \gg V$.

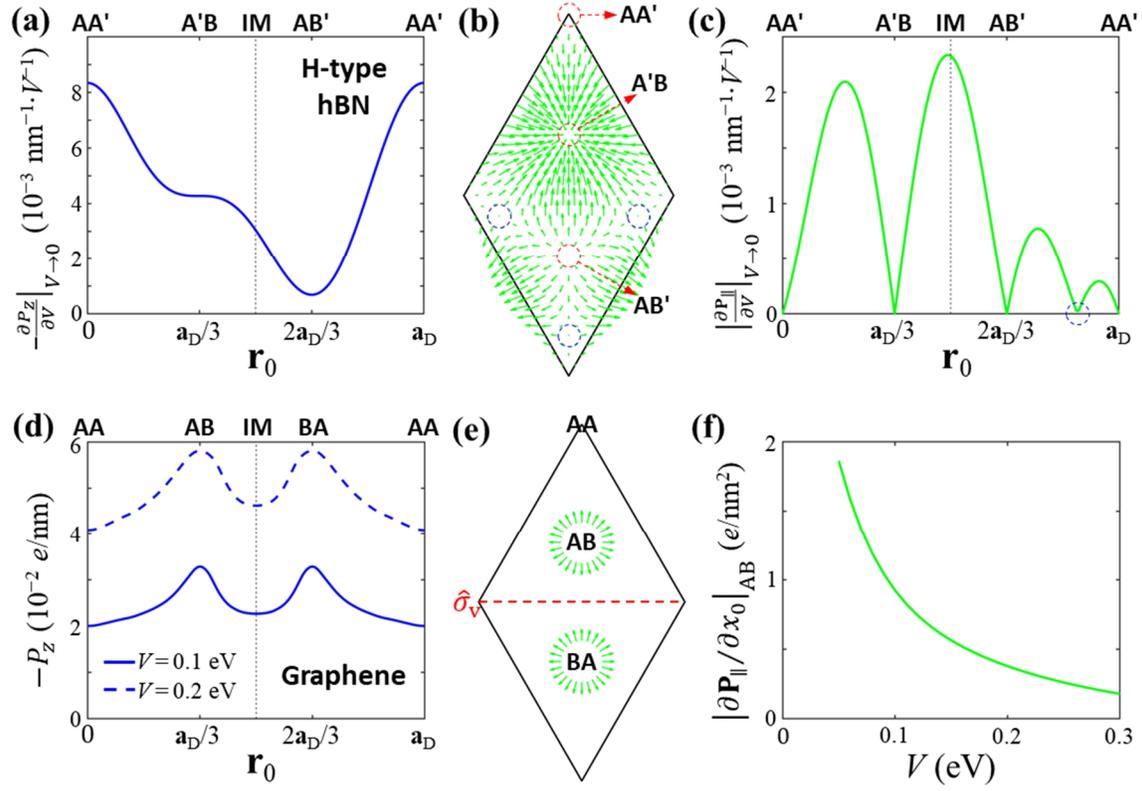

Fig. 3 (a) The calculated $\left.\frac{\partial P_z}{\partial V}\right|_{V\to 0}$ for $\mathbf{r}_0$ along $\mathbf{a}_D$ in the H-type bilayer hBN. (b) The direction of $\left.\frac{\partial \mathbf{P}_\parallel}{\partial V}\right|_{V\to 0}$ as a function of $\mathbf{r}_0$. The three red dashes circles correspond to AA', A'B and AB' stackings with +1 winding numbers. The three blue dashes circles correspond to three low-symmetry stackings with −1 winding numbers. (c) The magnitude $\left|\frac{\partial \mathbf{P}_\parallel}{\partial V}\right|_{V\to 0}$ for $\mathbf{r}_0$ along $\mathbf{a}_D$, which also vanishes at a low-symmetry stacking marked by the blue circle. (d) The calculated $P_z$ for $\mathbf{r}_0$ along $\mathbf{a}_D$ in bilayer graphene, under $V=0.1$ and 0.2 eV. (e) The directions of $\mathbf{P}_\parallel$ around AB/BA under $V>0$. The red dashed line indicates the vertical mirror plane ($\hat{\sigma}_v$) which transforms AB to BA and vice versa. (f) $\left|\frac{\partial \mathbf{P}_\parallel}{\partial x_0}\right|_{AB} = \left|\frac{\partial \mathbf{P}_\parallel}{\partial y_0}\right|_{AB}$ as a function of $V$.

It is natural to expect from the above formalism that a finite $\mathbf{P}_\parallel$ can also emerge under the condition $\mathsf{D}=\mathsf{D}'=0$, where the Hamiltonian describes a biased bilayer graphene. We set the parameters for graphene as $t=2.4$ eV, $t_{\mathbf{AA'}}^0 = t_{\mathbf{BB'}}^0 = t_{\mathbf{AB'}}^0 = t_{\mathbf{BA'}}^0 = 0.8$ eV, $s=0.15$ nm. Fig. 3(d) shows our calculated $P_z$ for $\mathbf{r}_0$ along $\mathbf{a}_D$ under $V=0.1$ and 0.2 eV, where a modulation of $\sim 0.01$ $e\cdot$nm$^{-1}$ with $\mathbf{r}_0$ has been obtained. For $\mathbf{P}_\parallel$, we focus on stackings which slightly deviate from AB/BA, i.e., $\mathbf{r}_0 = \pm \mathbf{a}_D/3 + d\mathbf{r}_0$, such that $\mathbf{P}_\parallel \approx \left.\frac{\partial \mathbf{P}_\parallel}{\partial x_0}\right|_{AB} dx_0 + \left.\frac{\partial \mathbf{P}_\parallel}{\partial y_0}\right|_{AB} dy_0$. Fig. 3(e) illustrates

the directions of $\mathbf{P}_\parallel$ around AB/BA under $V > 0$. In contrast to the R-type bilayer hBN shown in Fig. 2(d), in biased bilayer graphene the $\mathbf{P}_\parallel$ patterns are the same around AB and BA since they can be transformed to each other by a vertical mirror reflection ($\hat{s}_v$), see Fig. (3e). Upon changing the sign of $V$, both $P_z$ and $\mathbf{P}_\parallel$ reverse their directions. Fig. 3(f) indicates that the magnitude $\left|\frac{\partial \mathbf{P}_\parallel}{\partial x_0}\right|_{AB} = \left|\frac{\partial \mathbf{P}_\parallel}{\partial y_0}\right|_{AB}$ is in the order of 1 $e\cdot\text{nm}^{-2}$ for $V : 0.1$ eV. Thus, a tiny displacement $|d\mathbf{r}_0| : 0.1$ Å can introduce a sizable magnitude of $|\mathbf{P}_\parallel| : 0.01$ $e\cdot\text{nm}^{-1}$, the same order as $P_z$ in the R-type bilayer hBN. This implies that the shear phonon mode of the biased bilayer graphene can create an oscillating in-plane polarization. With the decrease of $V$, the gap between the conduction and valence bands gradually closes, resulting in the diverging behavior of $\left|\frac{\partial \mathbf{P}_\parallel}{\partial x_0}\right|_{AB}$. Note that the top valence and bottom conduction bands contribute opposite $\mathbf{P}_\parallel$ values, the overall $\mathbf{P}_\parallel$ thus averages to zero when the gap is much smaller than the temperature.

When the two layers have a small interlayer twisting and/or a tiny lattice mismatch, a long-wavelength moiré pattern will appear in the bilayer structure. In the moiré supercell, a local region (with a length scale small compared to the moiré wavelength) centered at $\mathbf{R}$ can be characterized by a commensurate stacking pattern with a position-dependent interlayer translation $\mathbf{r}_0(\mathbf{R})$. In the homobilayer system, the lattice mismatch can be induced by applying a heterostrain (strain difference between two layers). Below we consider three types of the formed moiré pattern in hBN bilayers, which originate from (1) an interlayer twisting $\theta$, (2) a biaxial-heterostrain $h_B$ and (3) an area-conserving uniaxial-heterostrain $\eta_U$ along the zigzag direction, respectively. All the three types realize hexagonal moiré superlattices with similar alternating patterns of AA, AB and BA locales, as illustrated in Fig. 4(a-c). However, IM locales between nearest-neighbor AB and BA have distinct crystalline orientations in the three cases. This comes from their different mapping functions between $\mathbf{R} = (X, Y)$ and $\mathbf{r}_0(\mathbf{R}) = (x_0(\mathbf{R}), y_0(\mathbf{R}))$, which up to the linear orders of $\theta$, $h_B$ and $\eta_U$ are given by [30]

$$\begin{cases} x_0(\mathbf{R}) = x_0(0) + qY, \\ y_0(\mathbf{R}) = y_0(0) - qX. \end{cases} \quad \text{(twisting } q\text{)}$$

$$\begin{cases} x_0(\mathbf{R}) = x_0(0) + h_B X, \\ y_0(\mathbf{R}) = y_0(0) + h_B Y. \end{cases} \quad \text{(biaxial-heterostrain } h_B\text{)} \quad (8)$$

$$\begin{cases} x_0(\mathbf{R}) = x_0(0) + h_U X, \\ y_0(\mathbf{R}) = y_0(0) - h_U Y. \end{cases} \quad \text{(area-conserving uniaxial-heterostrain } h_U\text{)}$$

In the above equation, the *xy*-coordinate directions are shown in Fig. 4(a-c). We did not take into account the lattice relaxation effect.

In a bilayer hBN moiré pattern with a twisting close to $0°$, the stacking registry dependent charge redistribution then gives rise to spatially varying electric polarizations with both out-of-plane and in-plane components. From Eq. (8) we get the electric polarization patterns as shown in Fig. 4(d-f). The three types of moiré patterns have the same spatial pattern for $P_z$. For $\mathbf{P}_\parallel$, however, although the spatial distributions of its magnitude are the same in the three cases, the directional patterns are sharply different, with distinct topologies. In twisting and biaxial-heterostrain moiré patterns with $\hat{C}_3$ symmetry, $\mathbf{P}_\parallel$ around AB/BA exhibits a winding number $w = +1$; whereas the corresponding winding number becomes $w = -1$ in the uniaxial-heterostrain case with broken $\hat{C}_3$ symmetry. $\mathbf{P}_\parallel$ in the twisting moiré corresponds to a divergence-free vector field with $\tilde{\nabla} \times \mathbf{P}_\parallel = 0$ and $\tilde{\nabla} \cdot \mathbf{P}_\parallel \neq 0$. In contrast, the biaxial-heterostrain moiré has $\tilde{\nabla} \times \mathbf{P}_\parallel \neq 0$ and $\tilde{\nabla} \cdot \mathbf{P}_\parallel = 0$, which implies a net lateral charge transfer that accumulates finite and opposite charges at AB and BA locales of the bilayer (Fig. 4(e)). Note that $\delta \rho_e = -\tilde{\nabla} \times \mathbf{P}_\parallel$ corresponds to the sum of the charge densities in the two layers. Assuming a moiré wavelength of 10 nm, the magnitude of $\delta \rho_e$ is estimated to be one order smaller than the charge density $P_z/d$ in the individual layer. For the uniaxial-heterostrain moiré, the $\mathbf{P}_\parallel$ vector field is both divergence- and curl-free, i.e., $\tilde{\nabla} \times \mathbf{P}_\parallel = 0$ and $\tilde{\nabla} \cdot \mathbf{P}_\parallel = 0$.

The different $\mathbf{P}_\parallel$ patterns of the three cases are rooted in their distinct symmetry properties. The twisting moiré pattern has three in-plane $\hat{C}_2$ axes related by the out-of-plane $\hat{C}_3$ rotation, which connect the nearest-neighbor AA locales (Fig. 4(d)). The vortex-like texture of $\mathbf{P}_\parallel$ is the result of these $\hat{C}_2$ axes combined with $\hat{C}_3$ symmetry. The biaxial-heterostrain moiré pattern has three vertical mirror planes ($\hat{\sigma}_v$), which are also related by $\hat{C}_3$ rotation. These $\hat{\sigma}_v$ planes correspond to the long diagonal lines of the diamond-shaped moiré supercells, resulting in the hedgehog-like texture of $\mathbf{P}_\parallel$ (Fig. 4(e)). The uniaxial-heterostrain moiré pattern, however, has only one $\hat{s}_v$ plane since $\hat{C}_3$ symmetry is broken (Fig. 4(f)). Taking into account both $P_z$ and $\mathbf{P}_\parallel$ components, the electric polarization pattern in each moiré supercell gives a pair of Bloch-type merons for the twisting case, Néel-type merons for the biaxial-heterostrain case, and anti-merons in the uniaxial-heterostrain case (note that here we define meron and anti-meron according to their winding numbers, not their topological charges [48]).

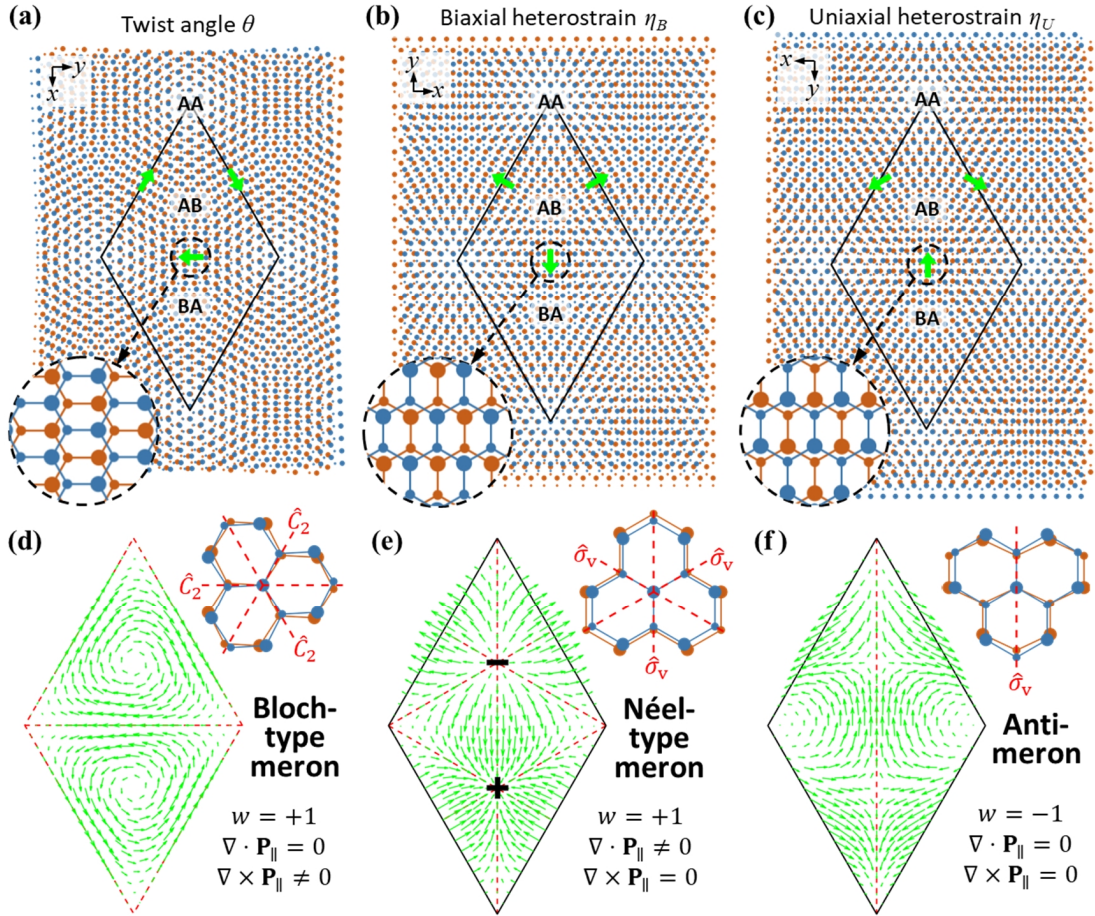

Fig. 4 Schematic illustrations of moiré patterns originating from (a) an interlayer twisting, (b) a biaxial-heterostrain and (c) an area-conserving uniaxial-heterostrain. The moiré supercells are indicated as diamond shapes, where alternating AA, AB and BA locales are marked. The atomic registries of IM locales between the nearest-neighbor AB and BA are enlarged, which show different polarization orientations in the three types of moiré patterns. (d-f) The corresponding $\mathbf{P}_\parallel$ directions as functions of location in a moiré supercell. The red dashed lines indicate the in-plane $\hat{C}_2$ axes or $\hat{s}_v$ planes of the three moiré patterns. The polarization patterns with distinct winding numbers, divergences and curls correspond to Bloch-type merons in the twisting case, Néel-type merons in the biaxial-heterostrain case, and anti-merons in the uniaxial-heterostrain

case. In the biaxial-heterostrain moiré pattern, the finite divergence results in opposite charge accumulations at AB and BA.

When both the twisting and heterostrain are present, both $\hat{C}_2$ and $\hat{s}_v$ symmetries are broken and the mapping functions between $\mathbf{R}$ and $\mathbf{r}_0(\mathbf{R})$ become

$$\begin{cases} x_0(\mathbf{R}) = x_0(0) + qY + h_B X, \\ y_0(\mathbf{R}) = y_0(0) - qX + h_B Y. \end{cases} \quad \text{(twisting + biaxial-heterostrain)}$$

$$\begin{cases} x_0(\mathbf{R}) = x_0(0) + qY + h_U X, \\ y_0(\mathbf{R}) = y_0(0) - qX - h_U Y. \end{cases} \quad \text{(twisting + uniaxial-heterostrain)} \tag{9}$$

For the 'twisting + biaxial-heterostrain' case, the resultant moiré pattern is always in a hexagonal form due to the presence of $\hat{C}_3$ symmetry. As shown in Fig. 5(a), the directions of $\mathbf{P}_\parallel$ around AB/BA are determined by $\varphi = \tan^{-1}(q/h_B)$, resulting in $\tilde{\nabla} \times \mathbf{P}_\parallel \neq 0$ and $\tilde{\nabla} \cdot \mathbf{P}_\parallel \neq 0$. When the values of $q$ and/or $h_B$ are continuously tuned, the winding number is always fixed at $w = +1$ and there is a smooth transition between the twisting moiré and biaxial-heterostrain moiré.

The 'twisting + uniaxial-heterostrain' case is completely different, where the formed moiré pattern is no longer hexagonal. In Fig. 5(b) and 5(c) we show two typical moiré patterns under $|q| > |h_U|$ and $|q| < |h_U|$, respectively, as well as their spatial patterns of $\mathbf{P}_\parallel$. The winding number for $|q| > |h_U|$ ($|q| < |h_U|$) is $w = +1$ ($-1$), the same as the twisting (uniaxial-heterostrain) moiré pattern. A topological transition happens at the mid-point $|q| = |h_U|$, which has the following mapping functions:

$$\begin{aligned} x_0(\mathbf{R}) - x_0(0) &= -(y_0(\mathbf{R}) - y_0(0)) = (Y + X)q, & (q = h_U) \\ x_0(\mathbf{R}) - x_0(0) &= y_0(\mathbf{R}) - y_0(0) = (Y - X)q, & (q = -h_U) \end{aligned} \tag{10}$$

A typical moiré pattern under $q = h_U$ is shown in Fig. 5(d). As can be seen, the local stacking registry doesn't change when moving $\mathbf{R}$ along $-45°$ direction, forming a moiré pattern with 1D modulation. When moving $\mathbf{R}$ along the registry-changing direction ($+45°$), its mapping function $\mathbf{r}_0(\mathbf{R})$ varies continuously along $-45°$ (Fig. 5(d) inset). We note that $\mathbf{r}_0(\mathbf{R})$ has no translational symmetry along $\pm 45°$ directions of the hexagonal lattice, so the 1D modulation of the local stacking registry and the resultant polarization are both quasiperiodic. In Fig. 5(e) we show $\mathbf{P}_\parallel$ and $P_z$ along the registry-changing direction in a length of $5L$, with $L \equiv a/|q|$ the characteristic length scale of the moiré pattern (i.e., the length of the thick line in Fig. 5(d)) and $a$ the monolayer lattice constant. Both $\mathbf{P}_{\parallel}$ and $P_z$ show strong modulations without periodicity. In fact, their Fourier transforms give three sets of 1D reciprocal lattice vectors which are incommensurate with each other (i.e., their ratios are all irrational numbers).

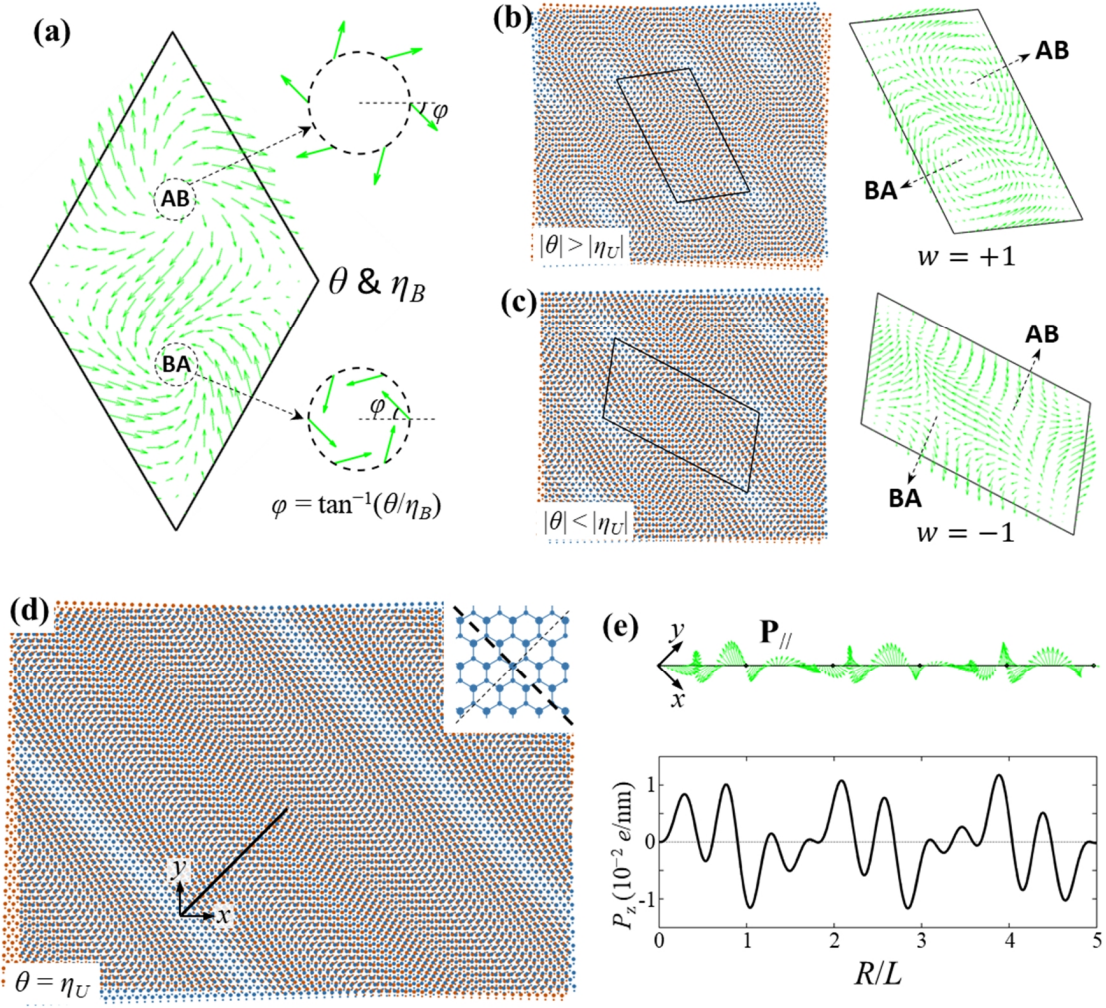

Fig. 5 (a) $\mathbf{P}_{\parallel}$ under the combination of twisting and biaxial-heterostrain. Around AB/BA locales, the directions of $\mathbf{P}_{\parallel}$ are determined by $q/h_B$. (b,c) Schematic illustrations of the non-hexagonal moiré patterns and the corresponding $\mathbf{P}_{\parallel}$ for the 'twisting + uniaxial-heterostrain' case, under $|q|>|h_U|$ and $|q|<|h_U|$, respectively. Each parallelogram corresponds to a moiré supercell. (d) A typical moiré pattern with 1D modulation under $q=h_U$, where we have set $\mathbf{r}_0(0)=0$. The solid thick line along the registry-changing direction (45°) corresponds to the characteristic length $L \approx a/|q|$. The dashed lines in the upper-right inset show the allowed values of $\mathbf{r}_0(\mathbf{R})$, where the thick (thin) line along $-45°$ (45°) is for $q=h_U$ ($q=-h_U$). $\mathbf{r}_0(\mathbf{R})$ has no translational symmetry along $\pm 45°$ directions of the hexagonal lattice. (e) The modulations of $\mathbf{P}_{\parallel}$ (upper) and $P_z$ (lower) along the registry-changing direction in (d).

In the uniaxial-heterostrain induced moiré pattern, the applied uniaxial strain can also give rise to an in-plane electric polarization in monolayer hBN and TMDs due to the piezoelectric effect, which corresponds to the first term on the right-hand-side of Eq. (3). From the calculated piezoelectric coefficient of ~ 0.8 $e$/nm of monolayer hBN [49], a 0.5% uniaxial strain can result in an in-plane polarization of $4\times10^{-3}$ $e$/nm, comparable to that originating from the interlayer coupling in Fig. 2(e). However, when the applied uniaxial strain is homogeneous, the induced polarization is also homogeneous and doesn't vary with position. We also note that in the above analysis to different moiré patterns, the two monolayers are treated as rigid and the lattice relaxation induced by the interlayer van der Waals interaction has been ignored. In realistic systems, the lattice relaxation can be significant when the moiré wavelength is large, such that the areas of the most stable registries AB and BA expand whereas the most unstable registry AA shrinks, and the mapping between $\mathbf{r}_0(\mathbf{R})$ and $\mathbf{R}$ is no longer a linear function [50]. This can result in inhomogeneous strain distributions with opposite signs in the two layers. In moiré patterns with close to $0°$ twisting, the in-plane polarizations induced by the inhomogeneous strains have opposite signs in the two layers which sum to zero [50]. In this case, the overall polarization is dominated by the interlayer coupling effect. On the other hand, in moiré patterns with close to $60°$ twisting, the strain induced polarizations have the same sign in the two layers thus could be more significant than those from the interlayer coupling. The local lattice relaxation can be further modulated by an external out-of-plane electric field due to its interaction with the interfacial electric dipoles, as illustrated in bilayer hBN and TMDs [34-38]. The IM locales with large in-plane polarizations thus will move under the effect of the out-of-plane electric field, which quantitatively changes the in-plane polarization in the long wavelength moiré pattern. The existence of $\mathbf{P}_\parallel$ component implies that an in-plane electric field can give rise to similar phenomena. Finally, we emphasize that the main features of the electric polarization patterns (including the Bloch-type merons for the twisting moiré, Néel-type merons for the biaxial-heterostrain moiré, anti-merons in the uniaxial-heterostrain moiré, their winding numbers and topological charges), which originate from the symmetries, are not changed by the lattice relaxation.

*Note added.* When finalizing the manuscript, we became aware of a related work [51], where in-plane electric polarizations in the R-type bilayer hBN are calculated from first-principles and discussed in the context of twisted and biaxially-strained hBN bilayers. Our results appear to agree where they overlap. We have shown that in-plane polarizations can also emerge in H-type hBN bilayer and graphene bilayer where interlayer bias breaks the inversion symmetry. Also we considered moiré patterns which originate from twisting, uniaxial-heterostrain, biaxial-heterostrain and their combinations, and show that they can all be distinguished by their contrasted spatial textures of in-plane polarizations.

*Acknowledgement*. H.Y. acknowledges support by NSFC under grant No. 12274477, and the Department of Science and Technology of Guangdong Province in China (2019QN01X061). W.Y. acknowledges support by the Research Grant Council of Hong Kong SAR (AoE/P-701/20, HKU SRFS2122-7S05), and Guangdong-Hong Kong Joint Laboratory of Quantum Matter.